\documentclass[%
 reprint,
 amsmath,amssymb,
 aps,
]{revtex4-2}

\usepackage{graphicx}% Include figure files
\usepackage{dcolumn}% Align table columns on decimal point
\usepackage{bm}% bold math
\usepackage{placeins}
\begin{document}

\preprint{APS/123-QED}

\title{Nanodroplet Condensation on Solid Surfaces}% Force line breaks with \\

\author{Matteo Teodori$^1$}
 \author{Dario Abbondanza$^1$}
 \author{Mirko Gallo$^{1}$} \email{mirko.gallo@uniroma1.it}
 \author{Carlo Massimo Casciola$^1$}
 
\affiliation{$^1$ Dipartimento di Ingegneria Meccanica e Aerospaziale, Sapienza Universit\`a di Roma,
Via Eudossiana 18, 00184 Roma {\it{Italy}}
}%

\date{\today}% It is always \today, today,

\begin{abstract}
This paper deals with the condensation of liquid droplets on hydrophobic and hydrophilic surfaces. A stochastic mesoscale model based on the theory of fluctuating hydrodynamics and the thermodynamics of a diffuse interface approach shows how direct simulation of the vapour-liquid transition from the nucleation process to droplet hydrodynamics can be achieved. Such simulations explain the role of wettability in filmwise and dropwise condensation regimes and the main limitations of classical nucleation theory.  

\end{abstract}

\maketitle

%\tableofcontents

\section{\label{sec:level1}Introduction}

Condensation is of central interest due to the innumerable conditions where vapour/liquid phase change takes place, from planetary -- e.g., clouds \cite{hoose2010important}-- to human -- e.g., water harvesting and desalination \cite{park2016condensation, luo2022advanced} -- and micro/nano scales -- e.g., biotechnology applications, electronics cooling \cite{wong2015micropatterned, grabowski2013growth, cho2015cool} and bionics/biomimetics \cite{ma2022condensation, shin2018hygrobot}. 
Fluid mechanics has long been involved in deciphering condensation \cite{ganapathy2013volume, orazzo2019direct, el2020review, rifert2020heat, katselas2022quantification}, an effort partially hindered by the poor understanding of the onset of the phenomenon,  namely the nucleation process \cite{lohse2016homogeneous}.

The classical picture of condensation is known. Vapour transforms into liquid by decreasing the temperature below the saturation point or by increasing the pressure above the liquefaction threshold. However, this description is oversimplified because it does not take into account metastability. 
In reality, a vapour may remain in a metastable state (local free-energy minimum) for a long time well beyond its thermodynamic saturation limits. Here the granular nature of matter plays a central role: molecules tend to aggregate with a certain probability resulting in nanometric droplets. A rare event consisting of a large enough cluster —a critical droplet—  requires a certain activation energy (nucleation barrier) to be exceeded, triggering the phase change. 
%Hence, to get quantitative information on phase change statistical mechanics must be invoked. 
Associated with the atomistic features of the process, nucleation also involves the macroscopic hydrodynamics of the droplets, namely the exchange of latent heat and coalescence.    

Macroscopic approaches currently do not account for thermal fluctuations and cannot describe droplet nucleation. 
These simulations are always initialized with pre-existing nuclei which are only partially consistent with the thermodynamics of phase change.  
%\cite{crialesi2023interaction} Citiamo chi traspoerta???
Consequently, it is impossible to predict whether the phase transformation begins within the vapour bulk (homogeneous nucleation) or in contact with a solid surface (heterogeneous nucleation) and whether droplets form in a film/dropwise mode. 
At the fundamental level, these different nucleation mechanisms are inherently connected with thermodynamic conditions and surface chemistry and proper modelling should take these aspects into account. 

%light to be shed on the two different condensation mechanisms observed: filmwise and dropwise condensation.  

Currently, Molecular Dynamics (MD) is the universally trusted tool for such a task \cite{ayuba2018kinetic,niu2018molecular,ranathunga2020molecular}, but its computational cost limits its use to nanometric systems on the nanosecond time scales. 
The coupling with macroscopic hydrodynamics became feasible by exploiting a recent mesoscale model originally developed for the reverse liquid/vapour transformation \cite{gallo_prf, gallo2018fluctuating, gallo2021heterogeneous, gallo2023nanoscale,gallo2024vapor}.

The methodology is based on a diffuse interface description \cite{van1893verhandel, van1979thermodynamic} combined with Fluctuating Hydrodynamics (FH) \cite{landau1980statistical}. FH is a powerful tool for assessing the influence of thermal fluctuations on macroscopic fluid dynamics \cite{chaudhri2014modeling, bandak2022dissipation, bell2022thermal, eyink2022high,barker2023fluctuating,eyink2024kraichnan}.  
%Numerical simulations provide the stochastic nucleation of liquid droplets in metastable vapours in both thermodynamic equilibrium and non-equilibrium (cold wall effect) contexts, allowing an unprecedented thermodynamic description of condensations within Navier Stokes hydrodynamics. 
Properly accounting for thermal fluctuations (Einstein-Boltzmann distribution) allows for the spontaneous nucleation of liquid droplets, describing condensation from nucleation to hydrodynamics. 
Using a prototypal fluid, we analyse the dependency on the solid surface chemistry of the different (dropwise/filmwise) condensation modes, when the system temperature is initially uniform.     

\section{\label{sec:level1}Mathematical Model} 
%Diffuse interface and fluctuating hydrodynamics simulations}

The proposed mesoscale model consists of the coupling between Landau-Lifshitz-Navier-Stokes equations and diffuse interface thermodynamics. Here the conservation equations for mass, momentum and energy are augmented with stochastic fluxes ($\delta\boldsymbol{\Sigma}$, $\delta\mathbf{q}$). 

\begin{equation}
\frac{\partial \rho}{\partial t} + \boldsymbol{\nabla} \cdot (\rho \boldsymbol{v}) = 0 \, , 
\label{mass}
\end{equation}

\begin{equation}
\frac{\partial \rho \boldsymbol{v}}{\partial t} + \boldsymbol{\nabla} \cdot (\rho \boldsymbol{v} 
\otimes  \boldsymbol{v}) = \boldsymbol{\nabla} \cdot \boldsymbol{\Sigma} + \delta\boldsymbol{\Sigma} \, , 
\label{momentum}
\end{equation}

\begin{equation}
\frac{\partial E}{\partial t} + \boldsymbol{\nabla} \cdot (\boldsymbol{v} E) = \boldsymbol{\nabla} \cdot [\left(\boldsymbol{\Sigma} +  \delta\boldsymbol{\Sigma}\right)\cdot \boldsymbol{v} - \boldsymbol{q}] + \delta\mathbf{q} \, , 
\label{energy}
\end{equation}
where $\rho$ is the mass density, $\boldsymbol{v}$ the velocity, 
and  $E$ the total energy density
 ($E=u_b(\rho,\theta)+1/2 \rho |v|^{2}+1/2\lambda |\boldsymbol{\nabla} \rho|^{2}$, with $u_b(\rho,\theta)$ the bulk internal energy density, 
 $\theta$ is the temperature, $\lambda$ the capillary coefficient, and $\boldsymbol{\nabla}$ the gradient operator). $\boldsymbol{\Sigma}$ and $\mathbf{q}$ are the deterministic stress tensor and energy flux as identified by non-equilibrium thermodynamics that include distributed capillary effects \cite{mcfadden, magaletti2016shock} (see also \cite{liu2016diffuse, benilov2023multicomponent, mukherjee2020understanding} for extensions to multi-component systems),  
 
 \begin{align*}
\boldsymbol{{\Sigma}}  =
\left[ -p + \frac{\lambda}{2} \vert\mathbf{\boldsymbol{\nabla}}\rho\vert^2 + \rho\mathbf{\boldsymbol{\nabla}}\cdot(\lambda\mathbf{\boldsymbol{\nabla}}\rho)\right] \mathbf{I} - \lambda\mathbf{\boldsymbol{\nabla}}\rho\otimes\mathbf{\boldsymbol{\nabla}}\rho +\\
+\eta_1(\mathbf{\boldsymbol{\nabla}}\mathbf{u} + \mathbf{\boldsymbol{\nabla}}\mathbf{u}^T)  + \eta_2 \mathbf{\boldsymbol{\nabla}}\cdot\mathbf{u}\mathbf{I} \, ,
\tag{4}\end{align*}

 \begin{align*}\label{eq:sigma-q}
\mathbf{q} = \lambda \rho \mathbf{\boldsymbol{\nabla}}\rho \mathbf{\boldsymbol{\nabla}}\cdot \mathbf{u} - k \mathbf{\boldsymbol{\nabla}} \theta \, ,
\tag{5}\end{align*}

where $p(\rho,\theta)$ is the thermodynamic pressure, $\mathbf{I}$ the identity matrix, the transport coefficients $\eta_1(\rho,\theta), \eta_2(\rho,\theta)$ are fluid's viscosities, and $k(\rho,\theta)$ the thermal conductivity. 

The fluctuation-dissipation balance provides the statistical properties of the Gaussian stochastic fluxes \cite{gallo2021heterogeneous}
\begin{align*}\label{eq:NS-system}
\nonumber
\langle \mathbf{\delta q}(\hat{\mathbf{x}},\hat{t})\rangle &=& 0 \, \qquad \langle \delta\boldsymbol{\Sigma}(\hat{\mathbf{x}},\hat{t})\rangle = 0 \, , \\
\nonumber
%\langle \mathbf{\delta\Sigma}(\hat{x},\hat{t})\rangle &=& 0 \, , \\
\nonumber 
\langle \delta\boldsymbol{\Sigma}(\hat{\mathbf{x}},\hat{t})\otimes \delta\boldsymbol{\Sigma}^\dagger(\tilde{\mathbf{x}},\tilde{t}) \rangle &=& \mathbf{Q^\Sigma}\delta(\hat{\mathbf{x}}-\tilde{\mathbf{x}}) \delta(\hat{t}-\tilde{t}) \, , \\ 
\langle \mathbf{\delta q}(\hat{\mathbf{x}},\hat{t}) \otimes\mathbf{\delta q}^\dagger(\tilde{\mathbf{x}},\tilde{t})\rangle &=& \mathbf{Q^q}\delta(\hat{\mathbf{x}}-\tilde{\mathbf{x}}) \delta(\hat{t}-\tilde{t}) \, , 
\tag{6}
\end{align*}
where $\delta(\mathbf{x})$ and $\delta(t)$ are the spatial and temporal Dirac delta functions, respectively, and 
\begin{align*}
%\nonumber
\mathbf{Q^\Sigma}_{\alpha\beta\nu\eta} = 2{\rm k}_B \theta \left[\eta_1( \delta_{\alpha\nu}\delta_{\beta\eta}+\delta_{\alpha\eta}\delta_{\beta\nu}) + \eta_2\delta_{\alpha\beta}\delta_{\nu\eta} \right]\,, \\
\nonumber
\mathbf{Q^q}_{\alpha\beta} = 2{\rm k}_B\theta^2 k\delta_{\alpha\beta} \, ,  
\tag{7}
\end{align*}
with the subscript Greek indices denoting the Cartesian components of tensors, ${\rm k}_B$ the Boltzmann constant, and $\delta_{\alpha\beta}$ the Kronecker delta symbols.    

To account for the fluid-solid interaction (i.e., the solid wettability), we have recently proposed a free-energy contribution \cite{gallo2021heterogeneous} which enforces the following boundary condition for the mass density, 

\begin{align}\label{eq:cadr}
\lambda \boldsymbol{\nabla} \rho \cdot \boldsymbol{\hat{n}}  = -\frac{df_w}{d\rho} = \cos\phi \sqrt{2 \lambda\left(\omega_b\left({\rho},\theta\right) - \omega_b\left(\rho_V\right)\right)}\, , 
\tag{8}\end{align} 

with  $\phi$ the Young contact angle, $\omega_b(\rho,\theta) = f_b -  \rho \partial f_b/\partial \rho$ and $f_b(\rho,\theta)$ the Landau and Helmholtz bulk free-energy densities, respectively, and $\hat{\mathbf{n}}$ the unit normal vector. 
It is worth stressing that the bulk energies $u_b, f_b, \omega_b$ and the pressure $p$ are identified by a proper equation of state (EoS). 
The EoS also defines the surface tension  \cite{benilov2020dependence, magaletti2021water}.  
In the present case, we adopt the modified Benedict-Webb-Rubin equation for mimicking a Lennard-Jones (LJ) fluid \cite{johnson1993lennard}. 

{\bf Numerical Setup:}
Condensation simulations are based on the system of Eq.s (\ref{mass}, \ref{momentum}, \ref{energy}). They are made dimensionless using the LJ reference quantities: $\sigma = 3.4 \times 10^{-10}$ m as length, $ \epsilon = 1.65 \times 10^{-21}$ J as energy, $m = 6.63 \times 10^{-26}$  kg as mass, $\theta_r = k_B/\epsilon = 119.56 \rm{K}$ as temperature, $v_r = (\epsilon/m)^{1/2} = 155.83  \rm{m/s}$ as velocity, and $t_r =  \sigma/v_r = 2.15\times10^{-12} \rm{s}$ as time.
To reproduce the surface tension of a LJ fluid and its temperature dependence, we set $\lambda = 5.224$, with $\lambda_r = (\sigma^5\epsilon)/m^2 = 1.52 \times10^{-18} \rm{kg m^7/s^2}$ \cite{gallo2022thermal}. 
Simulations consist of homogeneous metastable vapour enclosed in a box with two flat solid surfaces parallel to the $y-z$ plane. 
The boundary conditions on these surfaces are $\rho \boldsymbol{v} = 0$,  $\theta= \theta_w$, with the contact angle $\phi$ prescribed through Eq.(\ref{eq:cadr}),  and periodicity in $z$ and $y$ directions. 

The equations to be solved in a box of volume  $ V = 400 \times 1000 \times 1000$ are discretized on a uniform $40\times100\times100$ grid with the scheme proposed in \cite{balboa2012staggered} using a second-order Runge-Kutta scheme for time evolution ($\Delta t = 0.1$).  
This scheme was shown to preserve the fluctuation-dissipation balance at the discrete level \cite{donev2010accuracy, balboa2012staggered,delong2013temporal,magaletti2022positivity} and to 
guarantee converged statistics of thermal fluctuations \cite{gallo_prf, gallo2020nucleation, gallo2021heterogeneous}. 

\section{\label{sec:level1}Results and discussion}

%\subsubsection{Droplet condensation on solid surfaces with different wettability}

\begin{figure*}
	\centering
	{\includegraphics[width=1.7\columnwidth]{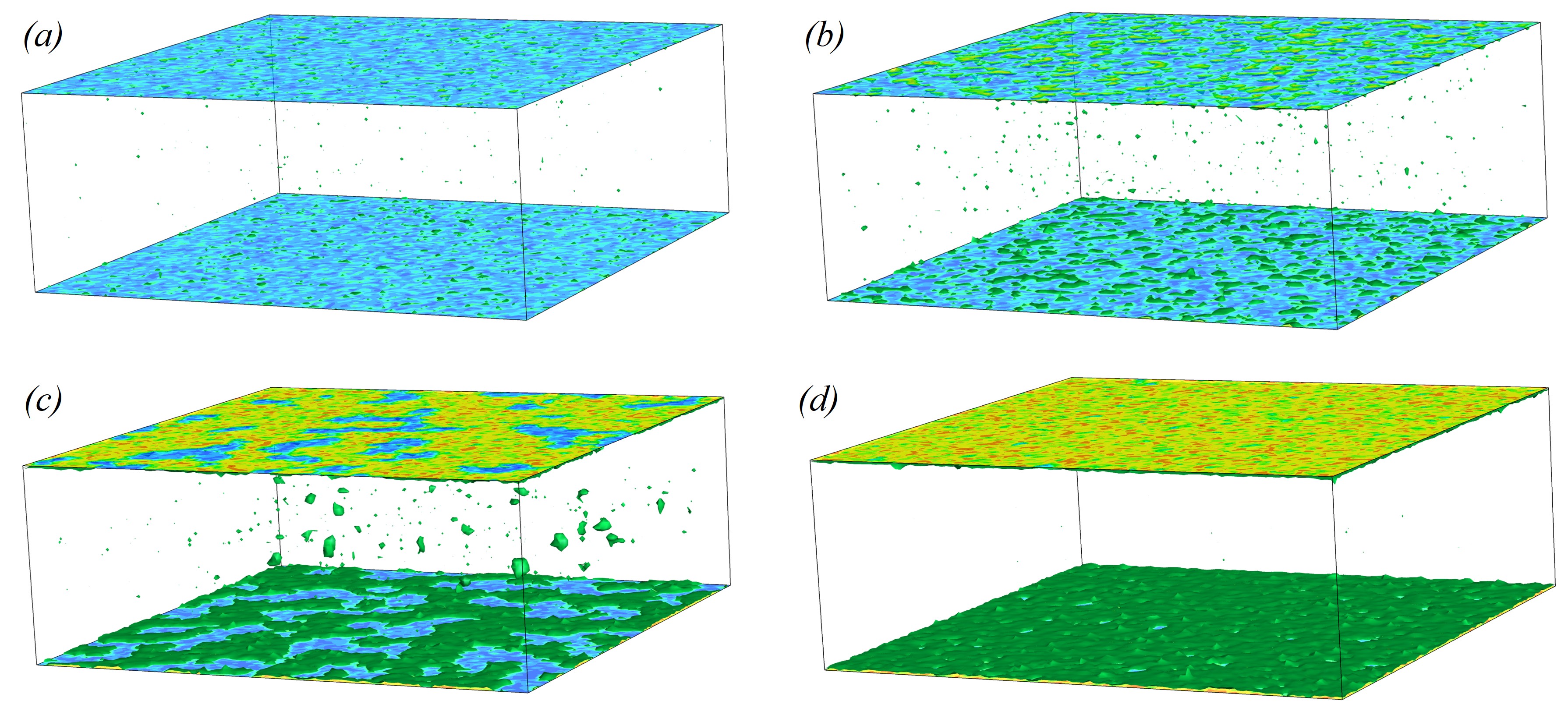}}\\
        {\includegraphics[width=1.5\columnwidth]{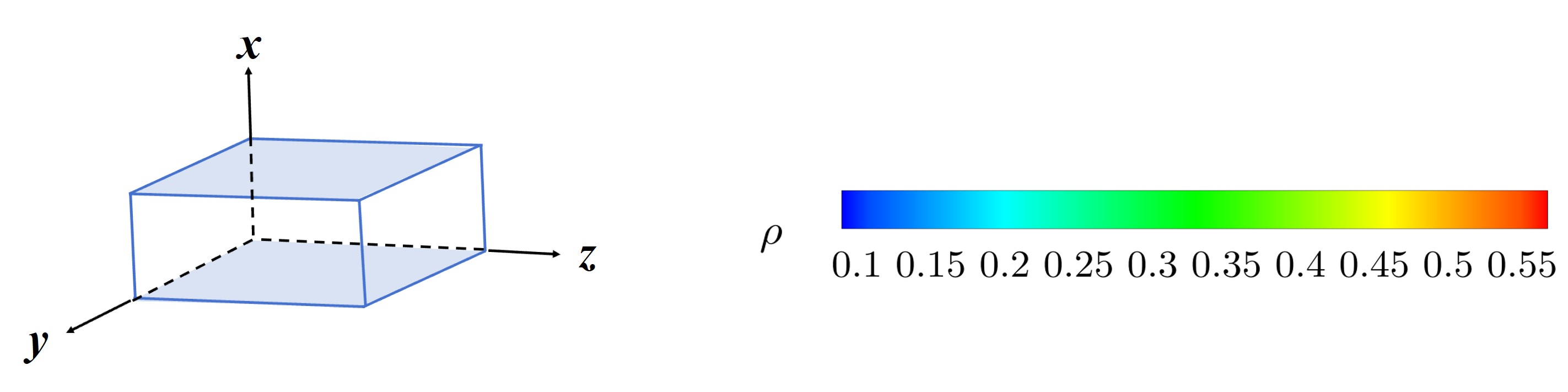}}\\
	\caption{Filmwise condensation: Snapshots of the density field at the solid walls, $\rho_{V}= 0.16$, $\phi = 30^{\circ}$. a) t = 600, b) t = 2000, c) t = 20000, and d) t = 150000. The isosurfaces at the critical density $\rho = 0.33$, taken to discern liquid from vapour, are shown in green. A sketch of the simulation box setup is also illustrated.}
	\label{1630}
\end{figure*}

The present simulation campaign investigates the dependence of condensation on the wettability of solid surfaces. 
The system is initialized with a uniform temperature $\theta = 1.25$ and an initial vapour density $\rho_{V}=0.16$, changing the contact angle in the range $15^{\circ}$-- 95$^{\circ}$.
In these conditions, the vapour is metastable, hence condensation is expected to occur.  
Snapshots of the density field along the simulation for $\phi=30^\circ$ are reported in Fig.~(\ref{1630}). 

Starting from a homogeneous metastable vapour, after a certain incubation time where sub-critical embryos are randomly formed and reabsorbed,  droplet nucleation begins to take place, Fig.~(\ref{1630}.a). 
% with the number of droplets linearly increasing with time, Fig.~(\ref{1630}.a) . 
In a successive phase, the number and size of droplets are sufficiently large to give rise to a number of coalescence events, Fig.~(\ref{1630}.b) and ~(\ref{1630}.c).
%the droplet number starts decreasing due to coalescence and reabsorption Fig.~(\ref{1630}.c). 
Eventually, a liquid film is formed as the precursor of the well-known film condensation mode in macroscopic experiments, Fig.~(\ref{1630}.d).
For $\rho_V = 0.16$, filmwise condensation is observed for contact angles below 60$^{\circ}$ where walls are rapidly covered by many nanodroplets that coalesce to form a continuous film.

\begin{figure*}
	\centering
	{\includegraphics[width=1.7\columnwidth]{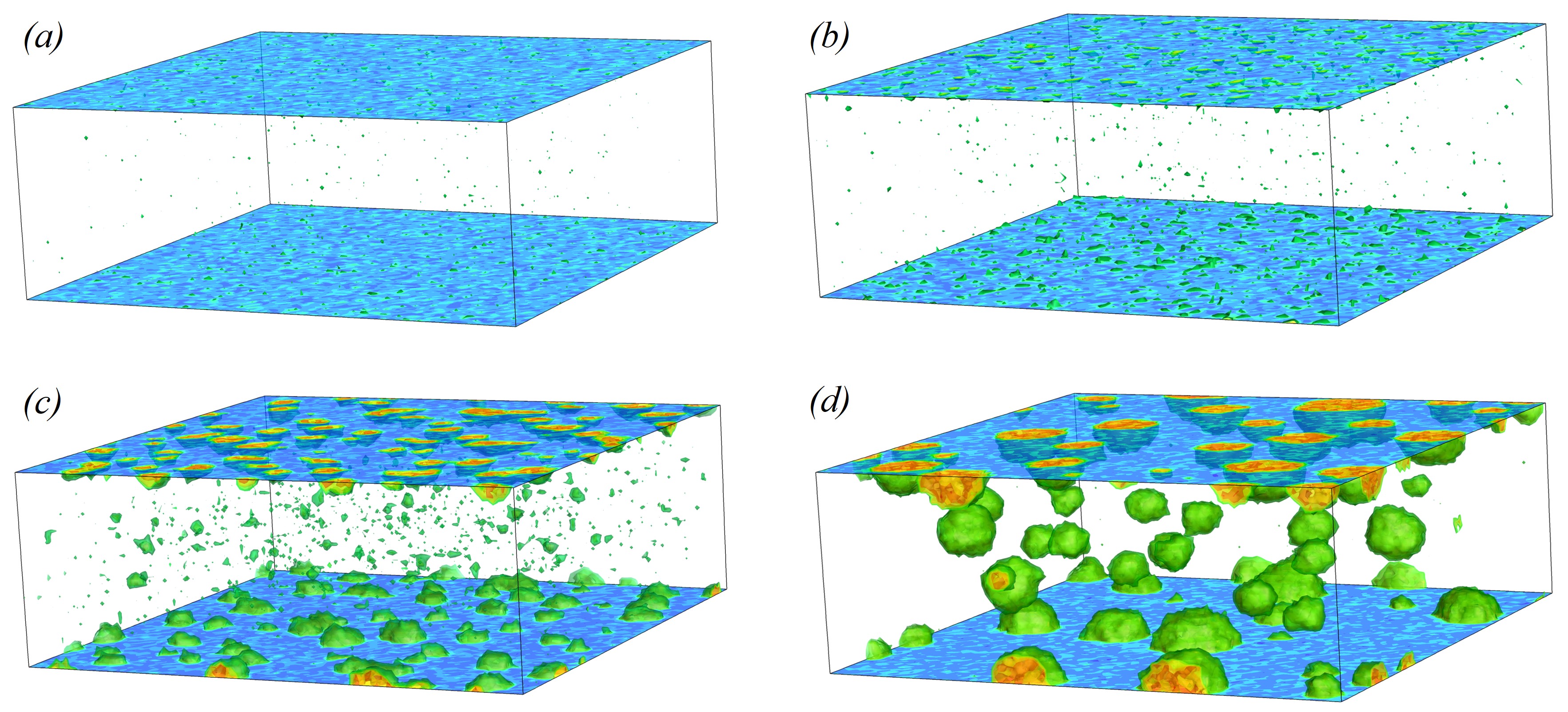}}\\
        {\includegraphics[width=1.5\columnwidth]{leg_ultima.jpg}}\\
	\caption{Dropwise condensation: Snapshots of numerical simulations for the case $\rho_{V}= 0.16$ and $\phi = 90^{\circ}$. a) t = 600, b) t = 2000, c) t = 20000, and d) t = 150000. The isosurfaces at $\rho = \rho_{cut} = 0.33$ that delimits the liquid phase are shown in green and the density field at the wall is reported. A sketch of the simulation box setup is also illustrated.} \label{1690}
\end{figure*}

At larger contact angles, the phenomenology is different. Only a few droplets nucleate, see Fig.~(\ref{1690}) where $\phi = 90^\circ$. After the
condensation nuclei appear, Fig.~(\ref{1690}.a), individual droplets form, Fig.~(\ref{1690}.b), that  
have enough time to grow as distinct entities, Fig.~(\ref{1690}.c), to occasionally coalesce, Fig.~(\ref{1690}.d).
Here we witness a sort of dropwise condensation mode at the nanoscale.

\begin{figure}
	\centering
	{\includegraphics[width=0.45\textwidth]{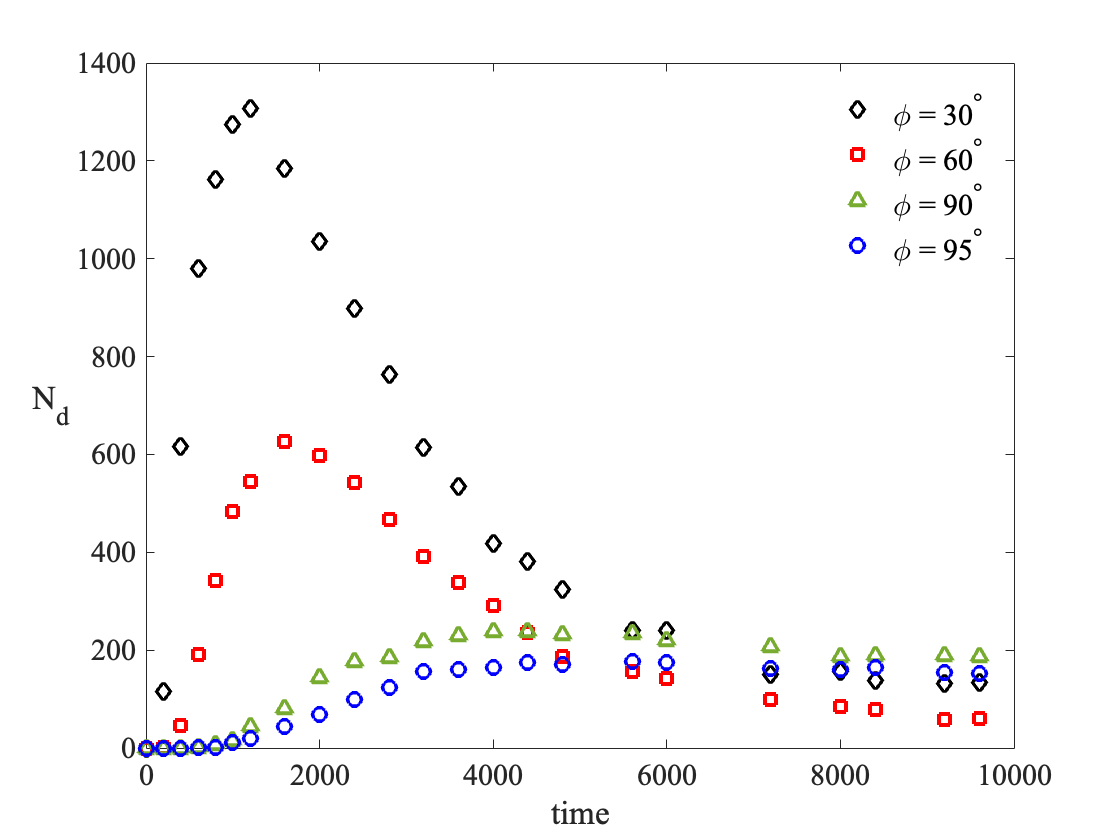}}\\
	\caption{Evolution of the number of supercritical droplets ($N_d$) over time. Different contact angles ($\phi$) are shown.}
 %The nucleation rate is related to the slope of the linear fit to the N$_{d}$ vs time curve in the nucleation phase of the condensation process. 
	\label{ndvstime}
\end{figure}

The evolution of the number of droplets over time, $N_d(t)$, at various wall wettabilities, is illustrated in Fig.(\ref{ndvstime}).
After the incubation time,  the number of droplets initially increases, linearly in time during a first phase.  Successively, the growth saturates and eventually, the number of detected drops steadily decreases. In part, some droplets are reabsorbed. This is due to the combined volume and total mass constraint which, after a certain amount of fluid liquefies, forces part of it to re-evaporate to occupy all the available space. This phenomenon is partly attributable to coalescence events which, while maintaining the overall volume of the liquid phase, lead to a decreased number of larger droplets. This picture is also observed in MD simulations \cite{kraska2006molecular}. 

Fig.~(\ref{ndvstime}) illustrates that the maximum number of formed droplets significantly decreases with increasing contact angle. Furthermore, following the peak, the subsequent decline in droplet count becomes more gradual. 

\begin{figure}[b]
		{\includegraphics[width=0.45\textwidth]{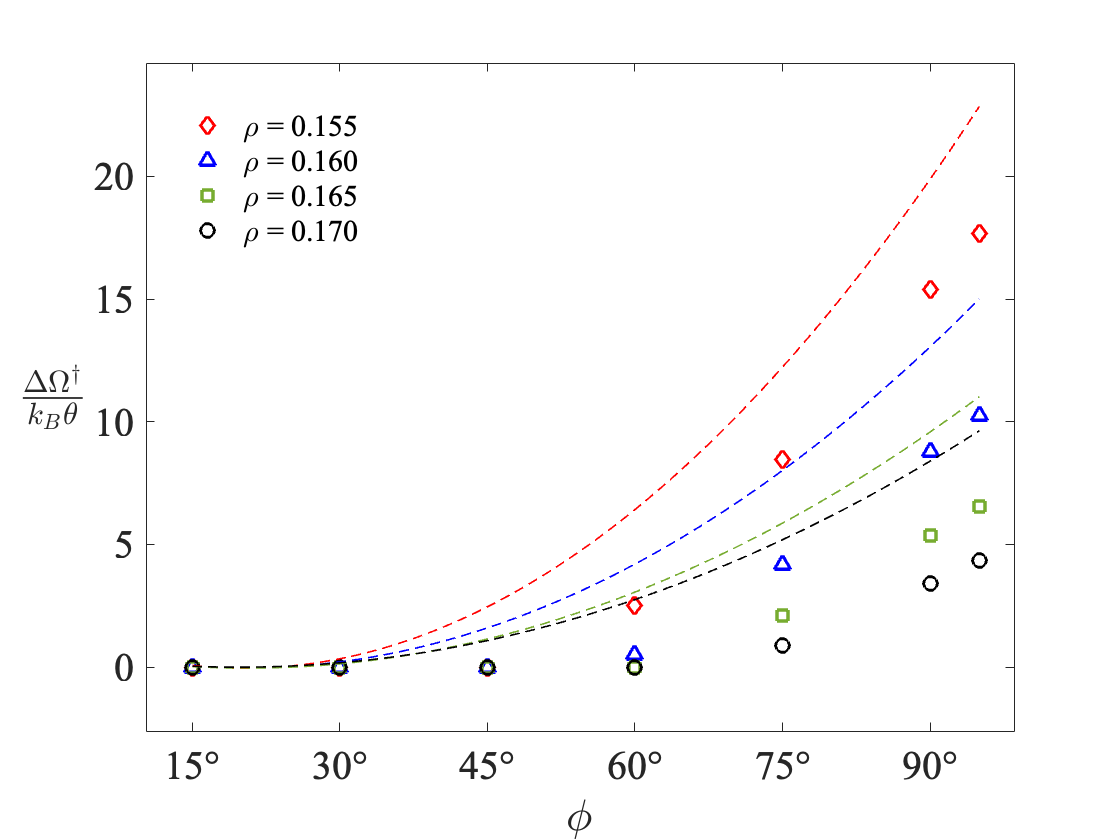}}\quad\
        {\includegraphics[width=0.45\textwidth]{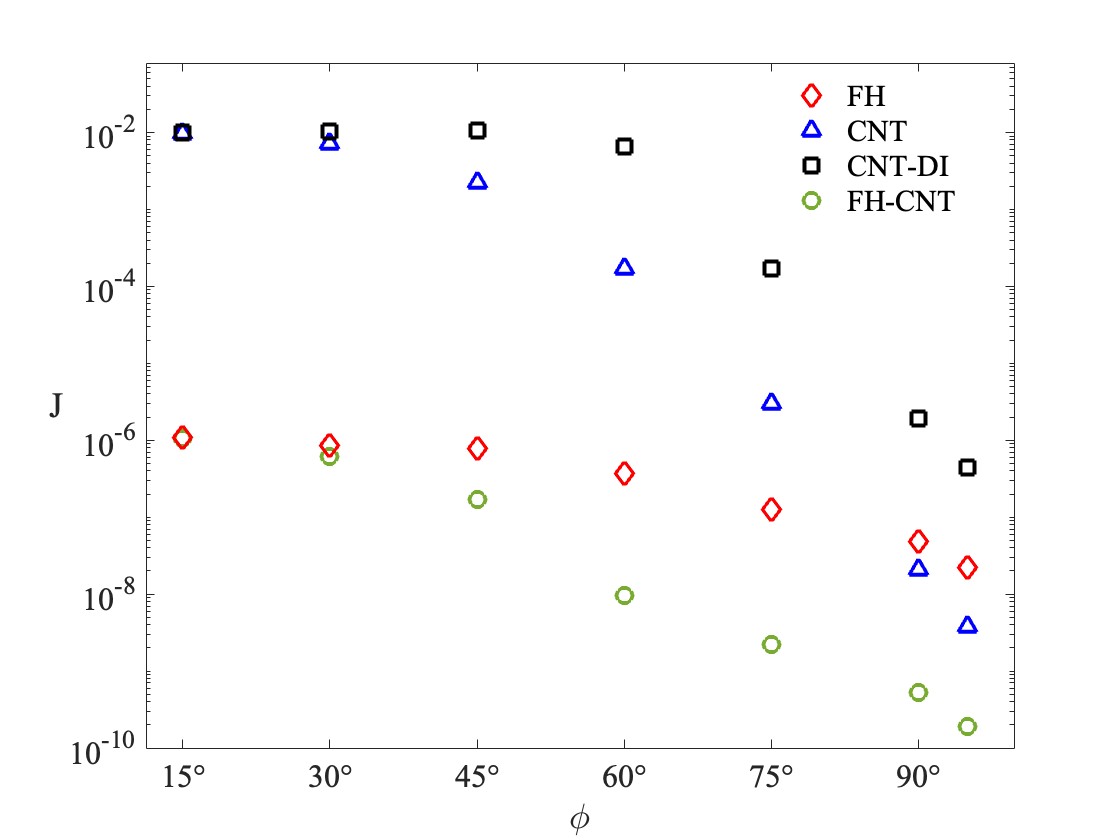}} 
	\caption{\textit{Top panel}: Normalised free energy barrier ($\Delta\Omega^\dag/k_B\theta$) vs contact angle ($\phi$).
    Dashed lines depict CNT, while symbols correspond to diffuse interface prediction. \textit{Bottom panel}: Nucleation rates ($J$) vs contact angle ($\phi$). Numerical simulations (red diamonds), CNT coupled with diffuse interface (CNT-DI black squares), CNT (blue triangles), FH coupled with CNT (green circles).}
	\label{jvsphi}
\end{figure}

In nucleation phenomena, the crucial observable is the nucleation rate $J$, related to the time history of the number of droplets, 
$N_d(t)$. In heterogeneous nucleation, it is defined as the number of droplets formed per unit time and unit wall area. In most MD simulations, where single droplet nucleation occurs due to the small spatial extent of the domain, the average time needed to observe a droplet is directly estimated from repeated simulations in identical thermodynamic conditions. For systems as large as the one currently addressed (fraction of micrometers), multiple nucleation events are observed in a single simulation. In this case, $J$ can be accurately determined from a single simulation \cite{yasuoka1998molecular}. As done in the context of cavitation \cite{gallo2021heterogeneous}, we adopt the same threshold method and measure $J$ from the slope of the linear fit to the N$_{d}$-vs-time curve in the (stationary) nucleation phase where $N_{d} \propto t$, i.e. $J= 1/A_w dN_d(t)/dt$, with $A_w$ the area of the condensing solid wall. Being Classical Nucleation Theory (CNT) the cornerstone of any discussion of (first-order) phase transition, it is natural to compare our estimates for the nucleation rates at different contact angles with those provided by CNT in the heterogeneous case $J_{\rm CNT} = n_V^{2/3}(n_V/n_L)(1-\cos(\phi))/2\sqrt{2\sigma/(\pi m\psi)}\exp(-\Delta\Omega^\dag/(k_B\theta))$, where $n_{V/L}$ is the number density of molecules in the vapour and liquid state, respectively, and $\psi = (1+cos(\phi))^{2}(2-cos(\phi))/4$ is a geometric factor that takes into account the contact angle \cite{kimura2002molecular}. 
The comparison shows that CNT gives an acceptable prediction for $\rho_{V} = 0.16$ only for contact angles near neutral wettability $\phi \simeq 90^\circ$. By decreasing the contact angle, the CNT is still qualitatively accurate, i.e. a higher nucleation rate at smaller contact angles, but the difference with the numerical results is not negligible (orders of magnitude). This is not dissimilar to what is observed in homogeneous and heterogeneous condensation when comparing Molecular Dynamics results with CNT (\cite{MD2013Diemand} and \cite{inci2011heterogeneous}, respectively). 
Now, we focus on identifying the primary source of error in the estimation of the nucleation rate, $J=J_0\exp{(-\Delta\Omega^\dag/(k_B\theta))}$, whether it originates from the pre-exponential factor $J_0$ or from the energy barrier $\Delta\Omega^\dag$ associated with the nucleation process. 
When dealing with nanodrops/bubbles, the typical dimension of the liquid/vapor interface is comparable with that of the droplet/bubbles, hence sharp-interface predictions --as CNT-- may fail \cite{menzl2016molecular,magaletti2021water}. Therefore, to precisely identify $\Delta\Omega^\dag$, the string method \cite{weinan2002string} is employed on top of diffuse interface thermodynamics ($\Delta\Omega^\dag_{DI}$). 
This is expected to be the barrier that the vapour phase experiences in our simulations. Fig.(\ref{jvsphi}) compares our numerical results with CNT and CNT-DI where the actual energy barrier is $\Delta\Omega^\dag_{DI}$ and $J_{CNT-DI}=J_{0}\exp(-\Delta\Omega^\dag_{DI}/(k_{B}\theta))$, with $J_{0} = n_V^{2/3}(n_V/n_L)(1-\cos(\phi))/2\sqrt{2\sigma/(\pi m\psi)}$ the usual CNT pre-exponential factor. 

\begin{figure}[hb!]
\hspace*{0pt}	{\includegraphics[width=0.45\textwidth]{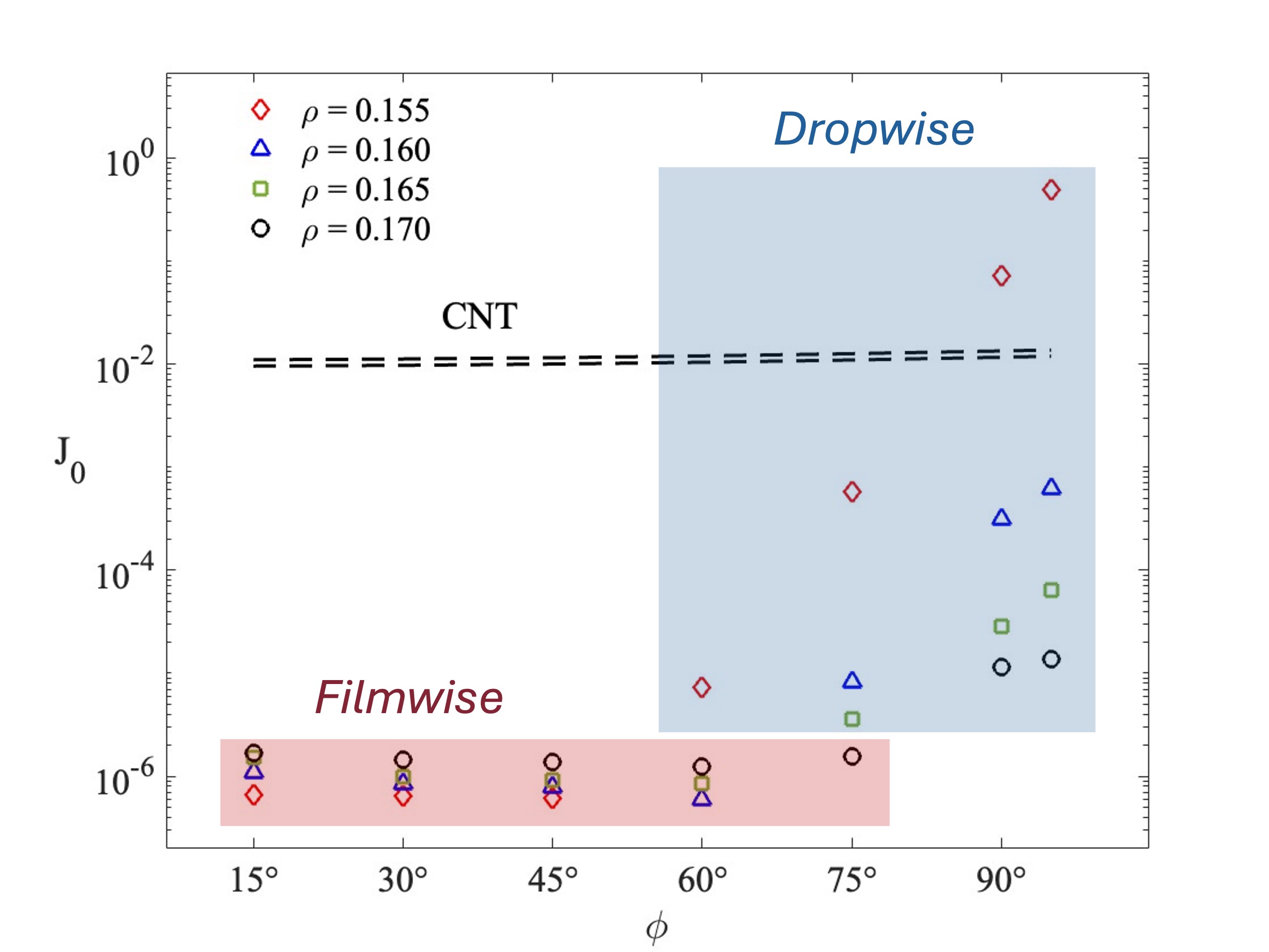}}
	\caption{Numerical values of $J_{0}(\rho,\phi)$ vs $\phi$ for different densities. The two dashed lines represent the minimum and maximum CNT prediction $J_0$ for each $\phi$ in the 0.155-0.17 density range.}
	\label{j0}
\end{figure}

As shown on the top panel of Fig.(\ref{jvsphi}), CNT overestimates the energy barrier over the entire range of observed contact angles at various degrees of metastability, in agreement with Molecular Dynamics simulations of heterogeneous condensation \cite{inci2011heterogeneous}. The nucleation rate $J$ in the two approaches overlaps only for low contact angles, where the barrier goes to zero and the nucleation rate is dominated by $J_{0}$, which is the same in both models. As the contact angle grows, the contribution of $J_{0}$ becomes less and less important and the nucleation rate heavily depends on the free-energy barrier.  It is also apparent that our numerical results, FH on the bottom panel of Fig.(\ref{jvsphi}), and the CNT-DI model are comparable only for more hydrophobic chemistry where the role of $J_0$ is not negligible. Hence, when the nucleation barriers are not so high, a significant part of the misalignment between CNT predictions and numerical data stems from a poor estimation of the factor $J_{0}$. From a physical perspective, the prefactor $J_0$
reflects both droplet diffusion in the vapor medium and the density of potential nucleation sites. In the homogeneous case, theoretical predictions \cite{langer1973hydrodynamic} and numerical simulations 
\cite{gallo_prf} suggest that FH offers a comprehensive description of both the energetic barrier ($\Delta\Omega^\dag$) and the kinetic aspects
($J_0$) of nucleation. Therefore, given a numerically obtained nucleation rate $J_{FH}$ we extract the corresponding FH pre-exponential factor as $J_{0_{FH}} = J_{FH}\exp{(\Delta\Omega_{ _{DI}}^\dagger/(k_B\theta))}$. Fig.(\ref{j0}) shows $J_{0_{FH}}$ evaluated at various $\phi$ for different initial vapour densities. For full context, we also report the nucleation rate resulting from the FH prefactor and CNT barrier, green circles in the bottom panel of Fig.(\ref{jvsphi}). In the latter case, $J$ is underestimated due to the CNT higher barriers.       

For comparison, the range of classical nucleation theory (CNT) prefactors is also reported (region between the two dashed lines). Two distinct trends, associated with different condensation regimes, are evident. In the filmwise condensation regime $J_{FH}$ remains nearly constant and is largely independent of the degree of metastability. In contrast, during dropwise condensation, the prefactor $J_0$	exhibits dependence on both $\phi$ and $\rho$. This behavior can be qualitatively interpreted using Langer’s theory \cite{langer1973hydrodynamic}, originally formulated for homogeneous nucleation. According to this framework, the prefactor scales with the system volume divided by the cube of the correlation length. In DI thermodynamics, the correlation length increases with metastability and diverges near the spinodal limit \cite{gallo2022thermal}. Consequently, the prefactor is expected to decrease as metastability increases—an effect not captured by CNT. Although Langer’s theory has not been formally extended to heterogeneous nucleation, its core concepts plausibly apply to dropwise condensation, which closely mirrors the dynamics of the homogeneous case.

\section{Conclusions}

Condensation is a complex phenomenon rooted in atomistic processes but also involving macroscopic hydrodynamic scales. For this reason, it has traditionally been studied at separate levels: nucleation through molecular dynamics simulations and bulk fluid behavior through Navier-Stokes-based solvers. In this work, we adopt a holistic approach, describing the entire process using a mesoscale model that combines hydrodynamic equations with the inherent granularity of matter treated as stochastic processes. This approach offers new insights into the various observed condensation regimes. Moreover, the relatively low computational cost ( $\sim 10^3$ core hours per simulation) paves the way for physically grounded, large-scale simulations of condensation in systems of significant technological relevance. 

\section*{Acknowledgements}
Funded by the European Union. Views and opinions expressed are however those of
the authors only and do not necessarily reflect those of the European Union or the
European Research Council Executive Agency. Neither the European Union nor the
granting authority can be held responsible for them. This work is supported by an ERC
grant (ERC-STG E-Nucl. Grant agreement ID: 101163330).
\section{Declaration of Interests} 
The authors report no conflict of interest.

\section*{Data availability}
The data that support the findings
of this study are openly available at \cite{teodori2025nanodroplets}.

%\nocite{*}
\bibliographystyle{apsrev4-2}
%\bibliography{mybibfile}% Produces the bibliography via BibTeX.


\end{document}